\documentclass[superscriptaddress,aps,amsfonts,notitlepage,twocolumn]{revtex4-2}
\usepackage{xcolor,graphicx}
\usepackage{breqn,amsmath,amssymb}

\usepackage{enumerate}

\usepackage{booktabs}
\usepackage{makecell}

\usepackage{comment}

\usepackage{subcaption}

\usepackage[T1]{fontenc}
\usepackage{natbib,hyperref}

\usepackage{mathtools}

\usepackage{centernot}

\makeatletter
\let\cat@comma@active\@empty
\makeatother

\usepackage{IEEEtrantools}

\begin{document}

\bstctlcite{IEEEexample:BSTcontrol}

\title{Reviving Horndeski after GW170817 by Kaluza-Klein compactifications}

\author{S. Mironov}
\email{sa.mironov\_1@physics.msu.ru}
\affiliation{Institute for Nuclear Research of the Russian Academy of Sciences, 
60th October Anniversary Prospect, 7a, 117312 Moscow, Russia}
\affiliation{Institute for Theoretical and Mathematical Physics,
MSU, 119991 Moscow, Russia}
\affiliation{NRC, "Kurchatov Institute", 123182, Moscow, Russia}

\author{A. Shtennikova}
\email{shtennikova@inr.ru}
\affiliation{Institute for Nuclear Research of the Russian Academy of Sciences, 
60th October Anniversary Prospect, 7a, 117312 Moscow, Russia}
\affiliation{Institute for Theoretical and Mathematical Physics,
MSU, 119991 Moscow, Russia}
\affiliation{Department of Particle Physics and Cosmology, Physics Faculty, M.V. Lomonosov Moscow State University,
Vorobjevy Gory, 119991 Moscow, Russia}

\author{M. Valencia-Villegas}
\email{mvalenciavillegas@itmp.msu.ru}
\affiliation{Institute for Theoretical and Mathematical Physics,
MSU, 119991 Moscow, Russia}

\begin{abstract}
The application of Horndeski theory/ Galileons for late time cosmology is heavily constrained by the strict coincidence in the speed of propagation of gravitational and electromagnetic waves. These constraints presuppose that the minimally coupled photon is not modified, {\it not even at the scales where General Relativity (GR) may need modification}. We find that the 4D Galileon obtained from a Kaluza-Klein compactification of its higher dimensional version is a natural {\it simultaneous modification} of GR and electromagnetism with  {\bf  automatically "luminal" gravitational waves}. This property follows without any fine tuning of Galileon potentials for a larger class of theories than previously thought. In particular, the $G_4$ potential is not constrained by the speed test and $G_5$ may also be present. In other words, some Galileon models that have been ruled out since the event GW170817 are, in fact, not necessarily constrained if they arise in 4D from compactifications of their higher dimensional Galileon counterparts. Besides their compelling luminality, the resulting {\bf vector Galileons} are naturally $U(1)$ gauge invariant. We also argue that the Vainshtein screening that allows to recover GR predictions for solar system tests is also at work for electrodynamics in the dense region of laboratory tests.
\end{abstract}

\maketitle

Galileon models and more generally Horndeski theory \cite{horndeski1974second,nicolis2009galileon,Deffayet:2011gz} - which is the most general modification of General Relativity (GR) by a scalar field with higher derivatives in the action, but with second order equations of motion - have played a major role in theoretical cosmology of the early and late time universe and for other interesting nonsingular solutions, because they may violate the Null Energy Condition without obvious pathologies \footnote{It is known, however, that in the general case, and without specific asymptotics \cite{Libanov:2016kfc, Kobayashi:2016xpl,Mironov:2019fop,Mironov:2022quk} or other constructions with Torsion \cite{Mironov:2023wxn,MM20242,Ahmedov:2023lot}, Horndeski theory has an issue with global instability. This issue will not be analyzed in this note.}. As general as they are, they include non-minimal couplings of the scalar with gravity that in most cases modifies the speed of gravity ($c_g$) with respect to the speed of light ($c=1$). Thus, it has been widely accepted that Horndeski theory is heavily constrained -at least for late time cosmology- by the measurement of the speed of gravitational waves (GW) since the event GW170817, which has set the bounds on the speed of gravity very close to the same of light, at $\vert \frac{c_g}{c} -1\vert \leq 5\times 10^{-16} $ \cite{LIGOScientific:2017vwq,Ezquiaga:2017ekz}.

It is clear that these constraints only apply to the original formulation of Horndeski theory with the photon minimally coupled to gravity and with speed $c=1$. In this letter we consider the idea that the Four Dimensional (4D) Galileon arises from a higher dimensional reduction, and thus the modified 4D graviton naturally comes alongside a modified photon. This simultaneous modification of GR and electromagnetism (EM) includes not only the usual non/minimal couplings of the  Galileon to the 4D graviton, but also new "nonminimal" couplings of the Galileon to the 4D photon. Thus, the speed of both tensor and vector modes is affected in similar ways -namely, the effective metric of both the graviton and photon is modified  by the same "gravito-electric" Galileon-, so as to preserve the same speed of propagation of GW and electromagnetic waves in 4D. As we will see, this enlarges the class of physically viable 4D Galileons that was singled out in \cite{Bettoni:2016mij,Ezquiaga:2017ekz}, {\it at least in what concerns the speed test}. Let us stress, however, that a dimensional reduction can be thought as merely
a tool. The essential aspect - the Galileon-photon couplings- can be obtained without higher dimensional ideas.

Now, higher dimensional scenarios leading to 4D Galileons are not new, and in fact present since their formulation. The initial Galileons were  inspired by the effective theory of the DGP braneworld model \cite{nicolis2009galileon} and some of their generalizations have been reproduced by {\it diagonal} \footnote{Namely, assuming that the 5D metric can be expressed in terms of our usual 4D metric and the dilaton, but without an extra vector field in the non-diagonal components.} Kaluza-Klein compactifications of Lovelock invariants \cite{VanAcoleyen:2011mj}, probe branes in higher dimensions \cite{deRham:2010eu,Trodden:2011xh} and from the low-energy effective action of heterotic string theory \cite{Easson:2020bgk} \footnote{See also \cite{Buchdahl:1979wi,Mueller-Hoissen:1987nvb} where Einstein-Maxwell Lagrangians of the form proposed in \cite{Horndeski:1976gi} were obtained by similar methods.}. In this letter, starting from a 5D gravity one naturally finds for every 4D-scalar Galileon a  4D-vector Galileon, such that {\bf the ratio} of the speed of light to gravity naturally remains a constant throughout the cosmic evolution, even with general potentials $G_4$ and $G_5$. The equations of motion remain of second order despite the higher derivatives of the vector and scalar in the action. The $U(1)$ gauge invariance is simply reconciled with the latter -as usual in a Kaluza-Klein compactification \cite{Kaluza:1921tu}, by diffeomorphism invariance in higher dimensions-, in contrast to previous constructions where eliminating ghosts or keeping gauge invariance is generally difficult \cite{Tasinato:2013oja,Petrov:2018xtx,Deffayet:2013tca,Colleaux:2023cqu,Colleaux:2024ndy}.

Thus, these theories are more robust against the speed test, potentially reviving previous efforts where non-minimal derivative couplings are essential (See for instance  \cite{Petronikolou:2021shp}), and yet, they can still be constrained with multi-messenger astronomy due to the very precise modifications to the vector sector. Furthermore, some of these theories naturally show nonminimal couplings of the vector to gravity, which are relevant for cosmological applications \cite{Esposito-Farese:2009wbc,Golovnev:2008cf,BeltranJimenez:2008iye}.

In section \ref{sec 1} we first show that in this formulation $G_4$ Galileons are not constrained by the speed tests of GW, then we extend this result to $G_5(\pi)$. In section \ref{sec screening} we discuss the Vainshtein mechanism for modifications of GR and electrodynamics. We argue that terrestrial experiments are safe due to the screening of modifications in both tensor and vector sectors near sources of gravity. In section \ref{sec conclusions} we give the conclusions.

\section{4D luminal vector-scalar Galileon from 5D gravity}\label{sec 1}
Let us start with the simplest Galileon model in 4D with nonminimal couplings to gravity: on top of GR take three scalar potentials $G_2,\, G_3,\, G_4$ that depend on $\pi$ and $X=-\frac{1}{2}\,g^{\mu \nu}\,\partial_\mu \pi\,\partial_\nu \pi$. The action
\begin{eqnarray}
\mathcal{S}_4&=&\int  \text{d}^4 x\,\sqrt{-g}\,\left(\mathcal{L}_2+\mathcal{L}_3+\mathcal{L}_4\right)\,,\nonumber\\
\mathcal{L}_2&=&G_2(\pi, \,X)\,,\nonumber\\
\mathcal{L}_3&=&G_3(\pi, \,X)\,\Box \pi\,,\nonumber\\
\mathcal{L}_4&=&\,G_4 \, R +G_{4,X}\,\left((\Box \pi)^2-(\nabla_\mu\nabla_\nu \pi)^2\right)\,,\label{eqn 4DGalileonsConstrained}
\end{eqnarray}
where $R$ is the Ricci scalar associated to the 4D metric and  $G_{4,X}=\partial G_4 / \partial X$, leads to second order equations of motion and allows nonsingular  solutions. Based on the bounds on the speed of GW, it is believed that the dependance of the $G_4$ potential on $X$ is heavily constrained. This can be explicitly seen in a first order perturbative expansion about a spatially flat FLRW background: with the perturbed metric $\textrm{d}s^2=\left(\eta_{\mu\nu}+\delta g_{\mu\nu}\right)\textrm{d}x^\mu\, \textrm{d}x^\nu $ where $\eta_{\mu\nu}\textrm{d}x^\mu\, \textrm{d}x^\nu= a^2(\eta)\left(-\textrm{d}\eta^2+\delta_{ij}\, \textrm{d}x^i \,\textrm{d}x^j \right)$, we consider only the symmetric, traceless and transverse tensor perturbation $h_{ij}$ and two transverse vector perturbations $S_i,\, F_i$, namely 
\begin{dmath}
\delta g_{\mu\nu}\,\textrm{d}x^\mu\, \textrm{d}x^\nu
\,=\, a^2(\eta)\left(2\,S_i \textrm{d}\eta \, \textrm{d}x^i+\left(\partial_i F_j+\partial_j F_i+2\,h_{ij}\right) \textrm{d}x^i \, \textrm{d}x^j \right)\,,
\end{dmath}
where $\eta$ is conformal time, and we denote spatial indices with latin letters such as $i=1,2,3$ and space-time indices with greek letters, such as $\mu=0,1,2,3$.  Furthermore, we write the perturbed Galileon field $\pi(x)$ as $\pi(\eta)+\chi(\eta,\vec{x})$ in the linearized expressions, where in this context $\pi(\eta) $ is the background field. With this notation, the quadratic action for the graviton derived from (\ref{eqn 4DGalileonsConstrained}) reads,
\begin{dmath}
\mathcal{S}_{4\,\tau} =\int\, \textrm{d}\eta\,\textrm{d}^3x \,a^4\,\left[\frac{1}{2\,a^2}\left({\mathcal{G}_\tau}\left(\dot{h}_{ij}\right)^2-{\mathcal{F}_\tau}(\partial_k\,{h}_{ij})^2\right)\right]\,,\label{eqn ql4tensor}
\end{dmath}
where $\mathcal{G}_\tau=2\,(G_4-2\,X\,G_{4,X}),\, \mathcal{F}_\tau=2\,G_4$ and the vector sector is nondynamical. Thus, for the speed of the tensor modes $c_g^2=\mathcal{F}_\tau/\mathcal{G}_\tau$ to exactly coincide with that of light, one is constrained to take only $G_4(\pi)$ potentials \cite{Bettoni:2016mij,Ezquiaga:2017ekz}. Similarly one finds that another possible potential in Horndeski ($G_5$), which contains nonminimal couplings of the scalar to gravity is totally ruled out if we assume a strict equality $c_g = c$ (See however \cite{Fernandes:2022zrq}).

This picture changes dramatically if we assume that the 4D Galileon arises from a higher dimensional reduction of a 5D theory. In this note we concentrate on the case that the 5D theory is itself a Galileon model (we do not discuss whether the 5D theory is fundamental or if it originates from yet higher dimensions, {\it e.g.} a higher dimensional Lovelock gravity \cite{VanAcoleyen:2011mj}). 

Let us consider the analogous $G_4$ Galileon model to (\ref{eqn 4DGalileonsConstrained}) but in $4+1$ dimensions,
\begin{eqnarray}
\mathcal{S}_5 =&\int & \text{d}^5 x\,\sqrt{-g_{(5)}}\, \Big(G_2(\pi,\,X)+G_3(\pi,\,X)\,\Box\pi\nonumber\\ 
&+& G_4\, R_{(5)}+G_{4,X}\,\left((\Box \pi)^2-(\nabla_A \nabla_B \pi)^2\right)\Big)\,,\label{eqn 5DGalileons}
\end{eqnarray}
where $R_{(5)}$ denotes the Ricci scalar computed with the five dimensional metric $g_{(5)}$. As usual, we can write the 5D metric as
\begin{equation}
g_{A\,B}=\left(
\begin{array}{cc}
g_{\mu\,\nu}+\phi^2\,A_\mu\,A_{\nu} & \phi^2\,A_\mu  \\  
\phi^2\,A_\nu & \phi^2
\end{array}\right)\,, \label{eqn kaluzametric}
\end{equation}
where $\phi$ is the dilaton and we will interpret $A_\mu$ as the EM Vector potential (See Section \ref{sec screening} for a discussion) \cite{Kaluza:1921tu}. We also follow the convention that uppercase latin letters are such as $B=0,\,1,\,2,\,3,\,$ or $4$ for the extra space dimension. Furthermore, for this note it is enough to take the Kaluza's cylinder condition and without further comment or more elegant dimensional reductions we assume that all physical quantities do not depend on the extra space dimension.

With the metric (\ref{eqn kaluzametric}), and the cylinder condition (normalizing our fields to reabsorb the multiplicative constant  $\int \text{d} x^4 $), we rewrite the 5D Galileon (\ref{eqn 5DGalileons}) as a 4D theory
\begin{eqnarray}
\mathcal{S}_5&=&\int \text{d}^4 x\,\sqrt{-g}\,\phi \, 
\bigg(\mathcal{L}_2+\mathcal{L}_3+\mathcal{L}_4+\mathcal{L}_{4\,\phi}+\phi^2\,\mathcal{L}_{4\,A}\bigg)\,\label{eqn 5DGalileonsIn4D}
\end{eqnarray}
where,
\begin{eqnarray}
\mathcal{L}_{4\,\phi}&=&- \frac{2}{\phi}\bigg(\,G_4 \,\Box \phi-\,G_{4,X}\, \, \nabla_\mu\phi \nabla^\mu \pi\,\Box\pi\bigg) \,, \\
\mathcal{L}_{4\,A}&=&-\,\bigg( G_4 \,\frac{1}{4}\,F^2 +\frac{1}{2}\,G_{4,X} \, F_\mu{}^\sigma\, F_{\nu\sigma}\,\nabla^\mu \pi\,\nabla^\nu \pi\bigg)\label{eqn L4A}
\end{eqnarray}
where $F_{\mu\nu}=\partial_\mu\,A_\nu-\partial_\nu\,A_\mu$ (Details are relegated to the Appendix \ref{secapp details}). As anticipated, this simultaneous modification of GR and EM in 4D is invariant under the gauge transformation $A_\mu\rightarrow A_\mu+\partial_\mu \Omega$, and it includes the usual nonminimal coupling of the Galileon to the graviton in $\mathcal{L}_4$, and the new "nonminimal" coupling of the Galileon also to the photon in $\mathcal{L}_{4\,A}$, Eqn. (\ref{eqn L4A}). The latter is essential so that, even with a general $G_4$ potential depending on $X$, the speed of the graviton and the photon is exactly the same, in contrast to (\ref{eqn 4DGalileonsConstrained}). Let us see: on the FLRW cosmological background let us consider the perturbed dilaton $\phi(x^\mu)$ as $\phi(\eta)+\Phi(x^\mu)$ and the perturbed photon $A_{\mu}$ written as $A_0(x^\mu)$ and $A_i(x^\mu)+\partial_i \alpha(x^\mu) $, where $\partial_i A_i=0$, and with vanishing vector background as we consider an isotropic spacetime. 

Clearly, the 4D {\it graviton} is nearly {\it the same as in the usual Horndeski theory}  (\ref{eqn ql4tensor}). There is again  only one tensor perturbation, it has the same speed, but now the Gravitational constant changes not only with the Galileon but also linearly with the dilaton, much in the spirit of Brans-Dicke (namely, taking $\mathcal{G}_\tau\rightarrow \phi\, \mathcal{G}_\tau$, $\mathcal{F}_\tau\rightarrow \phi\, \mathcal{F}_\tau$ in the Eqn. (\ref{eqn ql4tensor})). On the other hand, the quadratic action derived from (\ref{eqn 5DGalileonsIn4D}) for the vector perturbations can be written in momentum space as
\begin{dmath}
\mathcal{S}_{5\,V} =\frac{1}{2}\int\, \textrm{d}\eta\,\textrm{d}^3p \,a^2\,\phi\,\Big(\mathcal{G}_\tau\,\Big(\frac{1}{2\, a^2}\,\phi^2\,(\dot{A}_i)^2+\frac{1}{2}p^2\,(S_i)^2-p^2 S_i \dot{F}_i+\frac{1}{2}p^2 (\dot{F}_i)^2\Big)-\frac{1}{2\,a^2} p^2\, \mathcal{F}_\tau \phi^2 (A_i)^2 \Big) \,,\label{eqn ql5vector}
\end{dmath}
where we have used the equations for the background fields. Note that the photon decouples from the vector perturbations of the metric, which are nondynamical. Indeed, using the equation of motion for the metric perturbation $S_i=\dot{F}_i$, we reach the final quadratic action for the photon
\begin{dmath}
\mathcal{S}_{5\,V} =\frac{1}{4}\int\, \textrm{d}\eta\,\textrm{d}^3x \,\phi^3\,\left({\mathcal{G}_\tau}\left(\dot{A}_{i}\right)^2-{\mathcal{F}_\tau}(\partial_j\,{A}_{i})^2\right)\,,\label{eqn ql5vectorF}
\end{dmath}
and thus {\bf the speed of light $c^2=\mathcal{F}_\tau/\mathcal{G}_\tau$ coincides with the speed of GW for all $G_4$ potentials}. Let us stress that the truly essential aspect to the "luminality" is the completion of the Horndeski Lagrangian in Eqn. (\ref{eqn 4DGalileonsConstrained}) by the Galileon-Photon coupling in $\mathcal{L}_{4\,A}$, Eqn. (\ref{eqn L4A}). Neither the method how we obtained it -by a KK reduction- nor the dilaton are indispensable to this fact.

Notice that with the limit $A_\mu\rightarrow 0$ and $\phi\rightarrow$ constant in the action (\ref{eqn 5DGalileonsIn4D}), we recover the usual Horndeski action. Thus, (\ref{eqn L4A}) is a natural extension of Horndeski theory -but accompanied of a modified photon in 4D- which is more robust against speed tests of GW, and further analysis of multi-messenger phenomenology is necessary to constrain it.

Similarly, considering also the $G_5$ potential in the 5D Galileon (\ref{eqn 5DGalileons}), but restricting to the case that $G_5$ is only a function of $\pi$, but not of $X$,
\begin{dmath}
\int \text{d}^5 x\,\sqrt{-g_{(5)}}\, \Big( G_5(\pi)\,\left(R_{(5)}^{AB}-\frac{1}{2}g_{(5)}^{AB}\,R_{(5)}\right) \nabla_A \nabla_B \pi
\Big)\,,\label{eqn G5 5DGalileons}
\end{dmath}
we find again that {\it the speed of the graviton and photon in 4D coincide}.  This is in stark contrast with Horndeski theory directly formulated in 4D, where any nonzero $G_5$ leads to a difference in the speed of GW and the speed of light \cite{Bettoni:2016mij,Ezquiaga:2017ekz}. Explicitly, with the metric (\ref{eqn kaluzametric}) (See the Appendix \ref{secapp details}), the $G_5$ piece of the action (\ref{eqn G5 5DGalileons}) takes the form
\begin{eqnarray}
\int \text{d}^4 x\,\sqrt{-g}\,\,G_5(\pi)\,\left(\phi\,\mathcal{L}_{5}+\mathcal{L}_{5\, \phi} +\phi^3\, \mathcal{L}_{5\, A} \right)\,\label{eqn G5 5DGalileonsIn4D}
\end{eqnarray}
where,
\begin{eqnarray}
\mathcal{L}_{5}&=& \left(R^{\mu\nu}-\frac{1}{2}g^{\mu\nu}\,R\right) \nabla_\mu \nabla_\nu \pi\,,
\end{eqnarray}
\begin{eqnarray}
&&\mathcal{L}_{5\, \phi}= -\frac{1}{2}R\,\nabla_\mu \phi\,\nabla^\mu \pi +\left(\Box\phi\,\Box\pi-\nabla_\mu\nabla_\nu \phi \nabla^\mu\nabla^\nu \pi\right) \nonumber\\
&&+ \frac{3}{8}\,\phi^2 F^{\mu}{}_{\nu}\,F^{\sigma\nu}(-4\,g_{\lambda\mu}\,g_{\beta\sigma}+g_{\lambda\beta}\,g_{\mu\sigma})\,\nabla^\lambda \pi \nabla^\beta \phi \,,
\end{eqnarray}
\begin{eqnarray}
\mathcal{L}_{5\, A}&=&  \frac{1}{8}\, F^{\mu\nu}\,F_{\mu}{}^{\rho}\, (-4\,\nabla_\nu\nabla_\rho \pi +g_{\rho\nu}\Box\pi)\nonumber\\
&+&\frac{1}{2}\,F_{\mu\nu}\,\nabla_\sigma F^{\nu\sigma}\,\nabla^\mu \pi\,,\label{eqn L5A}
\end{eqnarray}
where $\mathcal{L}_{5\, A}$ contains all contributions to the photon, not mixed with derivatives of the dilaton. As before, {\it only} the new Galileon-Photon coupling in $\mathcal{L}_{5\, A}$ Eqn. (\ref{eqn L5A}) is {\it essential} for the luminality in Horndeski theory with $\mathcal{L}_{5}$. Let us note that the action  (\ref{eqn G5 5DGalileonsIn4D}) is a vector-scalar Galileon, namely, with higher derivatives of the scalar, the metric, as well as the photon in the action, but with second order equations of motion. Again, the $U(1)$ gauge invariance of the vector in 4D is ensured by the diffeomorphism invariance in 5D.

Finally, it is worth to notice that in 5D there is yet one more possible contribution to the Galileon action besides the $G_2$ to $G_5$ potentials of 4D Galileons \cite{Deffayet:2011gz}. We can write it with yet another scalar potential $G_6(\pi,X)$ and it takes the form 
\begin{eqnarray}
\int \text{d}^5 x\,\sqrt{-g_{(5)}}\,&\bigg(& G_6\,\mathcal{O}\Big(R_{(5)}{}^2\Big) +G_{6,X}\, \mathcal{O}\Big(R_{(5)}\,(\nabla^2 \pi)^2\Big) \nonumber\\
&+&G_{6,XX}\, \mathcal{O}\Big((\nabla^2 \pi)^4\Big)\bigg)\,,\label{eqn G6order}
\end{eqnarray}
where $\mathcal{O}$ denotes some polynomials of the given order on the Riemann curvature and second derivatives of $\pi$ ({\it e.g.} $\mathcal{O}\Big(R_{(5)}{}^2\Big)$ is the Gauss-Bonnet term). Similar as before, with the metric (\ref{eqn kaluzametric}), the action (\ref{eqn G6order}) is a (multi)scalar-vector Galileon in 4D, with the new aspect that now the photon is nonminimally coupled to gravity. {\it As with nonzero $G_{5,X}$, with a general $G_6$ potential, the speeds of photon and graviton do not coincide}. See however the analysis in Appendix \ref{secapp}.

\section{On the Vainshtein Screening of modifications near sources}\label{sec screening}

The physical feasibility of Horndeski modifications of General Relativity {\it crucially} relies on the Vainshtein effect \cite{Vainshtein:1972sx} (See also \cite{Koyama:2013paa,Babichev:2013usa,Kobayashi:2019hrl} for a review). It has been shown that in dense regions or the vicinity of a source -such as the solar system- the nonlinear self interactions of the scalar field in Eqn. (\ref{eqn 4DGalileonsConstrained}) become dominant, thus, effectively freezing the extra scalar mode $\pi$. Therefore, in the vicinity of a source we recover the usual two polarizations of the graviton and the predictions of GR. This is essential to pass the precise experimental constraints of solar system tests.

Let us see that the same screening mechanism may also be at work for Horndeski with the nonminimal scalar coupling to the photon $(\pi$-$A)$ in $\mathcal{L}_{4\,A}$ (\ref{eqn 5DGalileonsIn4D}). In the case of a constant dilaton $\phi$, in the vicinity of a source the scalar mode $\pi$ is frozen and {\it the effective description} of the dynamical modes of gravity can be accounted for with 
\begin{equation}
\begin{array}{cc}
G_4\rightarrow \text{constant}\,,& \mathcal{L}_{4}\rightarrow R
\end{array}\label{eqn GRlimit}
\end{equation}
and with the minimally coupled matter of $\mathcal{L}_{2}$ (namely, no $\mathcal{L}_{3},\, \mathcal{L}_{5}$). Thus, we note that {\it if} the Vainshtein screening of modifications of GR in {\it e.g.} the solar system is at work, then, with the limit (\ref{eqn GRlimit}) in Eqn. (\ref{eqn L4A}), in regions where $\partial^2 \pi$ becomes the dominant scale, we could also expect the standard Maxwell Electrodynamics at leading order near sources of gravity, 
\begin{equation}
\mathcal{L}_{4\,A}\rightarrow - \,\frac{1}{4}\,F^2\,.\label{eqn EMlimit}
\end{equation}
Indeed, building an effective action to analyze the Vainshtein screening, immediately reveals that the new $\mathcal{L}_{4\,A}$ in Eqn. (\ref{eqn 5DGalileonsIn4D}) {\it does not contribute to the second order derivative self-interactions of the scalar that are responsible for the screening of the extra mode}. For concreteness consider the simple case of a static, spherically symmetric configuration about a Minkowski background and with {\it constant} scalar and 4-vector backgrounds (vanishing electric and magnetic background fields). With $h$, $\chi$ and $A$ perturbations, respectively. Following \cite{Koyama:2013paa}, assuming that these perturbations are small, we neglect higher order interactions of the graviton, the vector and the scalar or its first-derivative self-interactions $(\partial \chi)^n$. We only keep nonlinearities of the form $(\partial^2 \chi)^n$ \cite{Koyama:2013paa,Kobayashi:2019hrl}. Furthermore, there are no linear terms in $\partial A$ of the form $\partial A\, (\partial^2\chi)^n  $ or $h\, \partial A$  (because $\mathcal{L}_{4\,A}$ is quadratic in $\partial A$ and the background 4-vector is constant). Finally, because the scalar background is also constant, any contribution {\it from the nonminimal terms} $F_\mu{}^\sigma\, F_{\nu\sigma}\,\nabla^\mu \pi\,\nabla^\nu \pi$ in $\mathcal{L}_{4A}$ Eqn. (\ref{eqn L4A}) to the linearized equation of motion for the Photon  {\it vanishes}. All in all, the terms that contribute to the effective theory take the form
\begin{eqnarray}
(\partial h)^2,\,(\partial \chi)^2,\,(\partial A)^2,\,  (\partial \chi)^2(\partial^2\chi)^n,\,h(\partial^2\chi)^n\,.\label{eqn effA}
\end{eqnarray}
Thus, with (\ref{eqn effA}) we reach the same effective action as in \cite{Koyama:2013paa}, but with the additional Maxwell term $(\partial A)^2$, and thus the conclusion on the screening of the extra scalar mode near sources, and (\ref{eqn GRlimit}), (\ref{eqn EMlimit}) follow. In short, we recover GR predictions and Maxwell electrodynamics in the dense region of laboratory tests, such that, for instance, {\it light does travel along geodesics near sources of gravity}.

However, this simple discussion merits more investigation about possible consequences and experimental constraints. Furthermore, the Vainshtein screening does not exclude the possibility of modifications to light traveling from distant sources, far from dense regions. Thus, Luminal extensions of Horndeski Gravity (\ref{eqn L4A}), (\ref{eqn L5A}) provide in principle experimentally distinguishable predictions, such as in cosmic rays.

\section{Conclusions}\label{sec conclusions}

We showed a new class of vector-scalar Galileons in 4D such that gravitational and electromagnetic waves naturally propagate at the same speed, even if one considers nonminimal derivative couplings of the scalar to gravity, thus, enlarging the class of physically viable 4D Galileons beyond those constrained by the event GW170817 \cite{Bettoni:2016mij,LIGOScientific:2017vwq,Ezquiaga:2017ekz,Abdalla:2022yfr} to also include the general $G_4$ and $G_5$ potentials of Horndeski theory. 

It was obtained from a Kaluza-Klein compactification from a 5D scalar Galileon. In particular, the equations of motion remain of second order despite the higher derivatives of the vector and scalar in the action. The $U(1)$ gauge invariance is guaranteed by  diffeomorphism invariance in higher dimensions. Furthermore, the limit of zero vector and constant dilaton recovers the usual Horndeski action (4D). Thus, the proposed theories with (\ref{eqn L4A}) and (\ref{eqn L5A}) are a natural extension of Horndeski theory, but accompanied of a modified photon in 4D, which are more robust against speed tests of GW, and thus further analysis of multi-messenger phenomenology is necessary to constrain them\footnote{For example, but not exclusively, gravitational (and even electromagnetic) wave decay constraints should be studied for such theories \cite{Creminelli:2018xsv}.}. We stressed that the Kaluza-Klein compactification (along with the accompanying dilaton) is merely
a tool, which naturally leads to the  phenomenologically desirable Photon-Galileon couplings of Eqns. (\ref{eqn L4A}) and (\ref{eqn L5A}), such that $c=c_g$ throughout the cosmic evolution. We did not claim uniqueness. Namely, there could exist more general modifications of GR and electrodynamics with a luminal graviton that in principle hold no relation to higher dimensional theories. Finally, we argued that the  Vainshtein mechanism may act in a similar way {\it screening modifications} of GR and Maxwell electrodynamics in the dense regions of laboratory tests,  such that, for instance, {\it light does travel along geodesics near sources of gravity}, and still, these Luminal extensions of Horndeski Gravity may provide experimentally distinguishable predictions, such as in cosmic rays.
\newpage

\section{Appendices}
\subsection{Details for the dimensional reduction of 5D Galileons}\label{secapp details}

To obtain (\ref{eqn 5DGalileonsIn4D}) and (\ref{eqn G5 5DGalileonsIn4D}) it is enough to express the 5D Ricci tensor $R^{(5)}_{AB}$ and $\nabla^{(5)}_A\nabla^{(5)}_B \pi$ (with $\nabla^{(5)}$ the connection associated to the 5D metric in Eqn. (\ref{eqn kaluzametric})) in terms of 4D fields, as

\begin{eqnarray}
&&R^{(5)}_{\mu\nu}=R_{\mu\nu}-\frac{1}{\phi}\nabla_\mu\nabla_\nu \phi-\frac{1}{2}\phi^2\, F_{\mu\rho}F_{\nu}{}^{\rho}+A_\mu \, A_\nu \, R^{(5)}_{44}\nonumber\\
&&+A_{\nu}\left(R^{(5)}_{4\mu}-A_{\mu}\,R^{(5)}_{44}\right)+A_{\mu}\left(R^{(5)}_{4\nu}-A_{\nu}\,R^{(5)}_{44}\right)\,, \end{eqnarray}
where,
\begin{eqnarray}
R^{(5)}_{44}&=&-\phi\,\nabla_\mu \nabla^\mu \phi +\frac{1}{4}\phi^4\,F_{\mu\rho}F^{\mu\rho}\,\\
R^{(5)}_{\mu 4}&=&\frac{3}{2}\phi\,F_{\mu \nu}\nabla^{\nu}\phi+\frac{1}{2}\phi^2 \,\nabla^\nu F_{\mu\nu}+A_{\mu}\,R^{(5)}_{44}\,
\end{eqnarray}
and
\begin{eqnarray}
&&\nabla^{(5)}_\mu\nabla^{(5)}_\nu \pi=\nabla_\mu \nabla_\nu \pi +\phi\,A_{\mu}\,A_{\nu}\nabla_\rho\phi\nabla^{\rho}\pi\,\nonumber\\
&&-\frac{1}{2}\phi^2\,A_{\mu}\, F_{\nu\rho}\,\nabla^{\rho}\pi -\frac{1}{2}\phi^2\,A_{\nu}\, F_{\mu\rho}\,\nabla^{\rho}\pi\,\\
&&\nabla^{(5)}_4\nabla^{(5)}_\mu \pi=(\phi \,A_{\mu}\nabla_\nu \phi-\frac{1}{2}\phi^2\,F_{\mu\nu})\nabla^\nu \pi\,\\
&&\nabla^{(5)}_4\nabla^{(5)}_4 \pi=\phi\,\nabla_\mu \phi\, \nabla^\mu \pi\,.
\end{eqnarray}
\subsection{Luminality with the most general $G_5(\pi,\,X)$ potential}\label{secapp}
In the case that $G_5$ depends on $X$ or $G_6$ is present, the speed of the graviton and the photon is different.  However, let us recall that Horndeski theory (4D) is written only with the four potentials $G_2$ to $G_5$. Thus,  the extra $G_6$ potential that results in 4D after compactification provides an extra freedom to build the vector Galileon in 4D that makes the Horndeski theory luminal. Although less elegant than the automatic luminality of the cases discussed before, where no fine tuning of the potentials is necessary, let us expand on this  possibility: with the most general $G_2$ to $G_5$ potentials depending on $\pi$ and $X$, the nonzero $G_{5X}$ leads to $c\neq c_g$. However, now $c$ and $c_g$ also depend on $G_6$ and thus, we can solve the equation $c=c_g$ for the dependent variable $G_6(\pi,\,X)$ in terms of the other independent potentials $G_2$ to $G_5$. In other words, we may find the 5D Galileon (with fine tuned $G_6$) that leads in 4D to a luminal vector-scalar Galileon with the most general unconstrained potentials $G_2$ to $G_5$. For instance, let us consider the case of a constant dilaton background $(\dot\phi(\eta)=0)$: let us redefine the general scalar  potentials $G_4(\pi,X),\,G_5(\pi,X)$ in terms of the free  potential $G_6(\pi,X)$ \cite{Charmousis:2011bf,Kobayashi:2011nu}
\begin{eqnarray}
&&G_4\rightarrow \tilde{G}_4-3\,X\,\frac{\partial^2}{\partial \pi^2} \bigg(\int\,\text{d}X\, \Big(\frac{G_6}{X}+2\,G_{6,X}\Big)-2\,G_6\bigg)\,,\nonumber \\
&&G_5\rightarrow \tilde{G}_5-3 \frac{\partial}{\partial \pi}\int\Big(\frac{G_6}{X}+2\,G_{6,X} \Big) \text{d}X\,,
\end{eqnarray}
then, in accordance with the theorem, the tensor sector reduces to the standard form (\ref{eqn ql4tensor}) with $\mathcal{G}_\tau$ and $\mathcal{F}_\tau$ depending only on $\tilde{G}_4(\pi,\,X)$, $\tilde{G}_5(\pi,\,X)$ and their derivatives, as usual in 4D Horndeski. However, the action for the photon still depends on $G_6(\pi,X)$ and its derivatives. It can be written (in conformal time) as
\begin{dmath}
\frac{1}{4}\int\, \textrm{d}\eta\,\textrm{d}^3x \,\phi^3\,\left({\mathcal{G}_V}\left(\dot{A}_{i}\right)^2-{\mathcal{F}_V}(\partial_j\,{A}_{i})^2\right)\,,
\end{dmath}
with
\begin{eqnarray}
\mathcal{G}_V&=&\mathcal{G}_\tau -2\, X\, H\, \frac{\dot{\pi}}{a} \tilde{G}_{5,X}\nonumber\\
&-&6\,H^2\,\left(G_6-4\,X\,\left(G_{6,X}+X\,G_{6,XX}\right)\right)\,,\\
\mathcal{F}_V&=&\mathcal{F}_\tau +6\,\bigg(\bigg(2\,X\,\frac{\dot{H}}{a}+H\,\frac{\dot{\pi}\,\ddot{\pi}}{\,a^3}\bigg)\,G_{6,X}\nonumber\\
&+&2\,X\,H\,\bigg(\frac{\dot{\pi}\,\ddot{\pi}}{\,a^3}-2\,X\,H\bigg)\,G_{6,XX}-\Big(H^2+\frac{\dot{H}}{a}\Big)\,G_6\nonumber\\
&+&H\,\frac{\dot{\pi}}{a} \bigg(-G_{6,\pi}+2\,X\,G_{6,X\pi}\bigg) \bigg)\,.
\end{eqnarray}
Thus, given the speed of the photon $c^2=\frac{\mathcal{F}_V}{\mathcal{G}_V}$, just consider the 5D theory with $G_6(\pi,X)$ that gives the 4D vector-scalar Galileon with constant ratio $\frac{c_g}{c}\equiv 1$.

\section*{Acknowledgement}

The work of S.M. is partly supported by the grant of the Foundation for the Advancement of Theoretical
Physics and Mathematics “BASIS" and by RFBR grant 21-51-46010.


\bibliographystyle{IEEEtran}
\bibliography{KK5to4DHorndeski}

\begin{thebibliography}{10}
\providecommand{\url}[1]{#1}
\csname url@samestyle\endcsname
\providecommand{\newblock}{\relax}
\providecommand{\bibinfo}[2]{#2}
\providecommand{\BIBentrySTDinterwordspacing}{\spaceskip=0pt\relax}
\providecommand{\BIBentryALTinterwordstretchfactor}{4}
\providecommand{\BIBentryALTinterwordspacing}{\spaceskip=\fontdimen2\font plus
\BIBentryALTinterwordstretchfactor\fontdimen3\font minus
  \fontdimen4\font\relax}
\providecommand{\BIBforeignlanguage}[2]{{%
\expandafter\ifx\csname l@#1\endcsname\relax
\typeout{** WARNING: IEEEtran.bst: No hyphenation pattern has been}%
\typeout{** loaded for the language `#1'. Using the pattern for}%
\typeout{** the default language instead.}%
\else
\language=\csname l@#1\endcsname
\fi
#2}}
\providecommand{\BIBdecl}{\relax}
\BIBdecl

\bibitem{horndeski1974second}
G.~W. Horndeski, ``Second-order scalar-tensor field equations in a
  four-dimensional space,'' \emph{International Journal of Theoretical
  Physics}, vol.~10, no.~6, pp. 363--384, 1974.

\bibitem{nicolis2009galileon}
A.~Nicolis, R.~Rattazzi, and E.~Trincherini, ``Galileon as a local modification
  of gravity,'' \emph{Physical Review D}, vol.~79, no.~6, p. 064036, 2009.

\bibitem{Deffayet:2011gz}
C.~Deffayet, X.~Gao, D.~A. Steer, and G.~Zahariade, ``{From k-essence to
  generalised Galileons},'' \emph{Phys. Rev. D}, vol.~84, p. 064039, 2011.

\bibitem{Note1}
It is known, however, that in the general case, and without specific
  asymptotics \cite {Libanov:2016kfc,
  Kobayashi:2016xpl,Mironov:2019fop,Mironov:2022quk} or other constructions
  with Torsion \cite {Mironov:2023wxn,MM20242,Ahmedov:2023lot}, Horndeski
  theory has an issue with global instability. This issue will not be analyzed
  in this note.

\bibitem{LIGOScientific:2017vwq}
B.~P. Abbott \emph{et~al.}, ``{GW170817: Observation of Gravitational Waves
  from a Binary Neutron Star Inspiral},'' \emph{Phys. Rev. Lett.}, vol. 119,
  no.~16, p. 161101, 2017.

\bibitem{Ezquiaga:2017ekz}
J.~M. Ezquiaga and M.~Zumalac\'arregui, ``{Dark Energy After GW170817: Dead
  Ends and the Road Ahead},'' \emph{Phys. Rev. Lett.}, vol. 119, no.~25, p.
  251304, 2017.

\bibitem{Bettoni:2016mij}
D.~Bettoni, J.~M. Ezquiaga, K.~Hinterbichler, and M.~Zumalac\'arregui, ``{Speed
  of Gravitational Waves and the Fate of Scalar-Tensor Gravity},'' \emph{Phys.
  Rev. D}, vol.~95, no.~8, p. 084029, 2017.

\bibitem{Note2}
Namely, assuming that the 5D metric can be expressed in terms of our usual 4D
  metric and the dilaton, but without an extra vector field in the non-diagonal
  components.

\bibitem{VanAcoleyen:2011mj}
K.~Van~Acoleyen and J.~Van~Doorsselaere, ``{Galileons from Lovelock actions},''
  \emph{Phys. Rev. D}, vol.~83, p. 084025, 2011.

\bibitem{deRham:2010eu}
C.~de~Rham and A.~J. Tolley, ``{DBI and the Galileon reunited},'' \emph{JCAP},
  vol.~05, p. 015, 2010.

\bibitem{Trodden:2011xh}
M.~Trodden and K.~Hinterbichler, ``{Generalizing Galileons},'' \emph{Class.
  Quant. Grav.}, vol.~28, p. 204003, 2011.

\bibitem{Easson:2020bgk}
D.~Easson, T.~Manton, M.~Parikh, and A.~Svesko, ``{The Stringy Origins of
  Galileons and their Novel Limit},'' \emph{JCAP}, vol.~05, p. 031, 2021.

\bibitem{Note3}
See also \cite {Buchdahl:1979wi,Mueller-Hoissen:1987nvb} where Einstein-Maxwell
  Lagrangians of the form proposed in \cite {Horndeski:1976gi} were obtained by
  similar methods.

\bibitem{Kaluza:1921tu}
T.~Kaluza, ``{Zum Unit\"atsproblem der Physik},'' \emph{Sitzungsber. Preuss.
  Akad. Wiss. Berlin (Math. Phys. )}, vol. 1921, pp. 966--972, 1921.

\bibitem{Tasinato:2013oja}
G.~Tasinato, K.~Koyama, and N.~Khosravi, ``{The role of vector fields in
  modified gravity scenarios},'' \emph{JCAP}, vol.~11, p. 037, 2013.

\bibitem{Petrov:2018xtx}
P.~Petrov, ``{Galileon-like vector fields},'' \emph{Phys. Rev. D}, vol. 100,
  no.~2, p. 025006, 2019.

\bibitem{Deffayet:2013tca}
C.~Deffayet, A.~E. G\"umr\"uk\c{c}\"uo\u{g}lu, S.~Mukohyama, and Y.~Wang, ``{A
  no-go theorem for generalized vector Galileons on flat spacetime},''
  \emph{JHEP}, vol.~04, p. 082, 2014.

\bibitem{Colleaux:2023cqu}
A.~Coll\'eaux, D.~Langlois, and K.~Noui, ``{Classification of generalised
  higher-order Einstein-Maxwell Lagrangians},'' \emph{JHEP}, vol.~03, p. 041,
  2024.

\bibitem{Colleaux:2024ndy}
A.~Coll\'eaux, D.~Langlois, and K.~Noui, ``{Degenerate Higher-Order Maxwell
  Theories in Flat Space-Time},'' 4 2024.

\bibitem{Petronikolou:2021shp}
M.~Petronikolou, S.~Basilakos, and E.~N. Saridakis, ``{Alleviating H0 tension
  in Horndeski gravity},'' \emph{Phys. Rev. D}, vol. 106, no.~12, p. 124051,
  2022.

\bibitem{Esposito-Farese:2009wbc}
G.~Esposito-Farese, C.~Pitrou, and J.-P. Uzan, ``{Vector theories in
  cosmology},'' \emph{Phys. Rev. D}, vol.~81, p. 063519, 2010.

\bibitem{Golovnev:2008cf}
A.~Golovnev, V.~Mukhanov, and V.~Vanchurin, ``{Vector Inflation},''
  \emph{JCAP}, vol.~06, p. 009, 2008.

\bibitem{BeltranJimenez:2008iye}
J.~Beltran~Jimenez and A.~L. Maroto, ``{A cosmic vector for dark energy},''
  \emph{Phys. Rev. D}, vol.~78, p. 063005, 2008.

\bibitem{Fernandes:2022zrq}
P.~G.~S. Fernandes, P.~Carrilho, T.~Clifton, and D.~J. Mulryne, ``{The 4D
  Einstein\textendash{}Gauss\textendash{}Bonnet theory of gravity: a review},''
  \emph{Class. Quant. Grav.}, vol.~39, no.~6, p. 063001, 2022.

\bibitem{Vainshtein:1972sx}
A.~I. Vainshtein, ``{To the problem of nonvanishing gravitation mass},''
  \emph{Phys. Lett. B}, vol.~39, pp. 393--394, 1972.

\bibitem{Koyama:2013paa}
K.~Koyama, G.~Niz, and G.~Tasinato, ``{Effective theory for the Vainshtein
  mechanism from the Horndeski action},'' \emph{Phys. Rev. D}, vol.~88, p.
  021502, 2013.

\bibitem{Babichev:2013usa}
E.~Babichev and C.~Deffayet, ``{An introduction to the Vainshtein mechanism},''
  \emph{Class. Quant. Grav.}, vol.~30, p. 184001, 2013.

\bibitem{Kobayashi:2019hrl}
T.~Kobayashi, ``{Horndeski theory and beyond: a review},'' \emph{Rept. Prog.
  Phys.}, vol.~82, no.~8, p. 086901, 2019.

\bibitem{Abdalla:2022yfr}
E.~Abdalla \emph{et~al.}, ``{Cosmology intertwined: A review of the particle
  physics, astrophysics, and cosmology associated with the cosmological
  tensions and anomalies},'' \emph{JHEAp}, vol.~34, pp. 49--211, 2022.

\bibitem{Note4}
For example, but not exclusively, gravitational (and even electromagnetic) wave
  decay constraints should be studied for such theories \cite
  {Creminelli:2018xsv}.

\bibitem{Charmousis:2011bf}
C.~Charmousis, E.~J. Copeland, A.~Padilla, and P.~M. Saffin, ``{General second
  order scalar-tensor theory, self tuning, and the Fab Four},'' \emph{Phys.
  Rev. Lett.}, vol. 108, p. 051101, 2012.

\bibitem{Kobayashi:2011nu}
T.~Kobayashi, M.~Yamaguchi, and J.~Yokoyama, ``{Generalized G-inflation:
  Inflation with the most general second-order field equations},'' \emph{Prog.
  Theor. Phys.}, vol. 126, pp. 511--529, 2011.

\bibitem{Libanov:2016kfc}
M.~Libanov, S.~Mironov, and V.~Rubakov, ``{Generalized Galileons: instabilities
  of bouncing and Genesis cosmologies and modified Genesis},'' \emph{JCAP},
  vol.~08, p. 037, 2016.

\bibitem{Kobayashi:2016xpl}
T.~Kobayashi, ``{Generic instabilities of nonsingular cosmologies in Horndeski
  theory: A no-go theorem},'' \emph{Phys. Rev. D}, vol.~94, no.~4, p. 043511,
  2016.

\bibitem{Mironov:2019fop}
S.~Mironov, ``{Mathematical Formulation of the No-Go Theorem in Horndeski
  Theory},'' \emph{Universe}, vol.~5, no.~2, p.~52, 2019.

\bibitem{Mironov:2022quk}
S.~Mironov and A.~Shtennikova, ``{Stable cosmological solutions in Horndeski
  theory},'' \emph{JCAP}, vol.~06, p. 037, 2023.

\bibitem{Mironov:2023wxn}
S.~Mironov and M.~Valencia-Villegas, ``{Stability of nonsingular cosmologies in
  Galileon models with torsion: A no-go theorem for eternal subluminality},''
  \emph{Phys. Rev. D}, vol. 109, no.~4, p. 044073, 2024.

\bibitem{MM20242}
S.~Mironov and M.~Valencia-Villegas, ``{Healthy Horndeski cosmologies with
  torsion},'' \emph{JCAP}, vol.~07, p. 030, 2024.

\bibitem{Ahmedov:2023lot}
B.~Ahmedov, K.~F. Dialektopoulos, J.~Levi~Said, A.~Nosirov, Z.~Oikonomopoulou,
  and O.~Yunusov, ``{Stable bouncing solutions in Teleparallel Horndeski
  gravity: violations of the no-go theorem},'' 11 2023.

\bibitem{Buchdahl:1979wi}
H.~A. Buchdahl, ``{ON A LAGRANGIAN FOR NONMINIMALLY COUPLED GRAVITATIONAL AND
  ELECTROMAGNETIC FIELDS},'' \emph{J. Phys. A}, vol.~12, pp. 1037--1043, 1979.

\bibitem{Mueller-Hoissen:1987nvb}
F.~Mueller-Hoissen, ``{Modification of Einstein {Yang-Mills} Theory From
  Dimensional Reduction of the {Gauss-Bonnet} Action},'' \emph{Class. Quant.
  Grav.}, vol.~5, p. L35, 1988.

\bibitem{Horndeski:1976gi}
G.~W. Horndeski, ``{Conservation of Charge and the Einstein-Maxwell Field
  Equations},'' \emph{J. Math. Phys.}, vol.~17, pp. 1980--1987, 1976.

\bibitem{Creminelli:2018xsv}
P.~Creminelli, M.~Lewandowski, G.~Tambalo, and F.~Vernizzi, ``{Gravitational
  Wave Decay into Dark Energy},'' \emph{JCAP}, vol.~12, p. 025, 2018.

\end{thebibliography}


\end{document}